# Large-scale comparison of bibliographic data sources: Scopus, Web of Science, Dimensions, Crossref, and Microsoft Academic

Martijn Visser, Nees Jan van Eck, and Ludo Waltman

Centre for Science and Technology Studies, Leiden University, The Netherlands

visser@cwts.leidenuniv.nl - https://orcid.org/0000-0001-5987-2389
ecknjpvan@cwts.leidenuniv.nl - https://orcid.org/0000-0001-8448-4521
waltmanlr@cwts.leidenuniv.nl - https://orcid.org/0000-0001-8249-1752

We present a large-scale comparison of five multidisciplinary bibliographic data sources: Scopus, Web of Science, Dimensions, Crossref, and Microsoft Academic. The comparison considers scientific documents from the period 2008–2017 covered by these data sources. Scopus is compared in a pairwise manner with each of the other data sources. We first analyze differences between the data sources in the coverage of documents, focusing for instance on differences over time, differences per document type, and differences per discipline. We then study differences in the completeness and accuracy of citation links. Based on our analysis, we discuss strengths and weaknesses of the different data sources. We emphasize the importance of combining a comprehensive coverage of the scientific literature with a flexible set of filters for making selections of the literature.

## 1. Introduction

Over the past 15 years, Web of Science (WoS; Birkle, Pendlebury, Schnell, & Adams, 2020; Schnell, 2017), Scopus (Baas, Schotten, Plume, Côté, & Karimi, 2020; Schotten, el Aisati, Meester, Steiginga, & Ross, 2017), and Google Scholar have been the three most important multidisciplinary bibliographic data sources, providing metadata on scientific documents and on citation links between these documents. It is very challenging to perform large-scale analyses using Google Scholar. WoS and Scopus have therefore long been the only options for large-scale bibliometric studies. This has changed in recent years with the introduction of two new multidisciplinary bibliographic data sources: Microsoft Academic (Sinha et al., 2015; Wang et al., 2019; Wang, Shen, Huang, Wu, Dong, & Kanakia, 2020) and Dimensions (Herzog, Hook, & Konkiel, 2020; Hook, Porter, & Herzog, 2018). At the same time, Crossref has become an increasingly valuable data source (Hendricks, Tkaczyk, Lin, & Feeney, 2020; see also Van Eck, Waltman, Larivière, & Sugimoto, 2018). Thanks to the Initiative for Open Citations (I4OC; https://i4oc.org/), launched in 2017, hundreds of millions of citation links between documents have been made openly available in Crossref. Likewise, the open availability of abstracts in Crossref is increasing thanks to the recently launched Initiative for Open Abstracts (I4OA; https://i4oa.org/).

Both for bibliometric research and for bibliometric practice, it is important to understand the strengths and weaknesses of different bibliographic data sources. Because most researchers do not have the possibility to retrieve large amounts of data from data sources such as Scopus and WoS, bibliographic data sources are typically compared in small-scale case studies, focusing for instance on documents in a specific research field or on a small number of researchers and the documents they have authored (e.g., Harzing, 2019). A large-scale comparison of Scopus and WoS was presented by Mongeon and Paul-Hus (2016). A similar comparison, including not only Scopus and WoS but also Dimensions, was carried out by Singh, Singh, Karmakar, Leta, and Mayr (2020). However, both comparisons were performed at the level of journals rather than individual documents. Recently, Huang et al. (2020) reported a document-level comparison of Scopus, WoS, and Microsoft Academic based on a fairly large amount of data (i.e., documents published by 15 universities). Their comparison has the limitation that documents in the different data sources are matched based only on Digital Object Identifiers (DOIs). Another recent document-level comparison was performed by Martín-Martín, Thelwall, Orduna-Malea, and López-Cózar (2020; see also Martín-Martín, Orduna-Malea, Thelwall, & López-Cózar, 2018). This comparison considers Scopus, WoS, Dimensions, OpenCitations, Microsoft Academic, and Google Scholar. The comparison starts by selecting a limited number of highly cited documents and then analyzes the overlap between the different data sources in terms of documents that cite the selected highly cited documents. The comparison involves over three million citing documents.

In this paper, we present a large-scale document-level comparison of five major bibliographic data sources: Scopus, WoS, Dimensions, Crossref, and Microsoft Academic. Our focus is on differences between the data sources in the coverage of documents. In addition, we also study differences in the completeness and accuracy of citation links. We consider only scientific documents, such as journal articles, preprints, conference proceedings papers, books, and book chapters, in our analysis. Some data sources also provide data on other types of entities. Dimensions for instance offers data on grants, data sets, clinical trials, patents, and policy documents. Likewise, the WoS platform provides data on data sets and patents. While this data can be of great value, it falls outside the scope of our analysis.

A number of major bibliographic data sources are not covered by the comparison presented in this paper. Google Scholar is not included in the comparison because we do not have large-scale access to this data source. Studies of Google Scholar typically focus on relatively small numbers of documents (e.g., Harzing, 2019; Martín-Martín, Orduna-Malea, & López-Cózar, 2018). Large-scale studies of Google Scholar (Martín-Martín, Thelwall et al., 2018; Martín-Martín et al., 2020) require an extraordinary amount of effort (Else, 2018). OpenCitations is another important data source that is not included in the comparison. It is not included because it currently provides more or less the same data as Crossref (Heibi, Peroni, & Shotton, 2019b). This is expected to change in the near future (Peroni & Shotton, 2020), so OpenCitations deserves careful attention in future work. Finally, the comparison does not cover PubMed. This data source is not included because it does not provide data on citation links between documents.

To keep the analysis manageable, we use Scopus as a baseline and we perform pairwise comparisons of Scopus with each of the other data sources. Because Scopus and WoS are the most established bibliographic data sources, it seems natural to use either of these data sources as the baseline in our analysis. We use Scopus rather than WoS as the baseline because we do not have access to the full WoS database. Our use of Scopus as the baseline does not mean that we consider Scopus to be our preferred bibliographic data source.

The rest of this paper is organized as follows. In Section 2, we provide a more detailed discussion of the data sources included in our analysis. The procedure developed for matching documents in different data sources is described in Section 3. We present the results of our analysis in Sections 4 and 5. Conclusions and limitations are discussed in Sections 6 and 7.

## 2. Data sources

In our analysis, we focus on scientific documents published in the period 2008–2017. Scientific documents can be articles in journals, but also preprints, papers in conference proceedings, books, book chapters, and so on. We consider the following five bibliographic data sources:

- *Scopus*. Scopus is a data source produced by Elsevier. Our center has full access to Scopus for documents starting from 1996. We use Scopus data delivered to our center in April 2019.
- *CWTS WoS*. WoS is a data source produced by Clarivate Analytics. Clarivate Analytics distinguishes between the WoS Core Collection and the broader WoS platform. Our focus is on the WoS Core Collection. The WoS Core Collection consists of a number of citation indices. We consider the Science Citation Index Expanded (SCIE), the Social Sciences Citation Index (SSCI), the Arts & Humanities Citation Index (AHCI), and the Conference Proceedings Citation Index (CPCI). Our center has full access to these citation indices for documents starting from 1980. The Emerging Sources Citation Index (ESCI) and the Book Citation Index (BKCI) are also part of the WoS Core Collection. We do not consider these citation indices, because our center does not have access to them. We use WoS data updated until the end of 2018. The data was delivered to our center in XML format. In the interpretation of our findings for WoS, it is essential to keep in mind that the Emerging Sources Citation Index and the Book Citation Index are not included in our analysis. In the rest of this paper, we use the label *CWTS WoS* to refer to the WoS data to which our center has access and to distinguish this data from the full WoS database.
- *Dimensions*. Dimensions is a data source produced by Digital Science. Our center has full access to Dimensions. We use Dimensions data delivered to our center in June 2019. In addition to scientific documents, Dimensions also covers grants, data sets, clinical trials, patents, and policy documents. We do not include this content in our analysis.
- *Crossref*. Crossref provides an infrastructure through which scientific publishers make metadata available for the content they publish. We use Crossref data downloaded in August 2018 through the public REST API of Crossref. We downloaded the data in JSON format. The following content types are excluded from our analysis: book-part, book-section, component, dataset, journal-issue, peer-review, posted-content, proceedings, proceedings-series, report-series, and standard.

- *Microsoft Academic*. Microsoft Academic is a data source produced by Microsoft. We use a dump of Microsoft Academic data from March 2019 (Microsoft Academic, 2019). Content classified as data set or patent is excluded from our analysis.

The different data sources have different content selection policies. WoS has an internal editorial team for content selection. WoS emphasizes the selectivity of its content selection policy for the WoS Core Collection, and in particular for the Science Citation Index Expanded, the Social Sciences Citation Index, and the Arts & Humanities Citation Index (Birkle et al., 2020; Schnell, 2017). Scopus works together with an international group of researchers, referred to as the Content Selection and Advisory Board, to perform content selection (Baas et al., 2020; Schotten et al., 2017). Scopus often emphasizes the size of its database. Compared with the WoS Core Collection, it therefore appears to focus more on comprehensiveness and less on selectivity. Dimensions has an even stronger focus on comprehensiveness: "The database should not be selective but rather should be open to encompassing all scholarly content that is available for inclusion … The community should then be able to choose the filter that they wish to apply to explore the data according to their use case." (Hook et al., 2018; see also Herzog et al., 2020). Microsoft Academic has the strongest focus on comprehensiveness. It claims to replicate "the success of Google Scholar, which utilizes the massive document index from a web search engine to achieve comprehensive coverage of contemporary scholarly materials, many of which are not published and distributed through traditional channels and not assigned DOIs" (Wang et al., 2020).

Crossref (Hendricks et al., 2020) is a special case. It is a registration agency for DOIs. When a scientific publisher works with Crossref to register a DOI for a document, the publisher provides metadata for this document to Crossref. This metadata is then made openly available by Crossref (with the possible exception of the reference list, for which the publisher determines whether it is made openly available or not). In this way, Crossref has become a bibliographic data source that is of significant interest for bibliometric analyses. The completeness and the quality of the data available in Crossref depend on what publishers provide to Crossref. Crossref itself does not actively collect and enrich data.

## 3. Matching of data sources

Because of the large amount of data, matching documents in Scopus with documents in the other data sources is a challenging task. We developed a matching procedure that aims to provide accurate results within an acceptable amount of computing time. This matching procedure is discussed in this section.

### 3.1. Preprocessing

Our matching procedure starts by preprocessing the data obtained from the different data sources. In the case of publication years and volume, issue, page, and article numbers, the preprocessing process retains only numerical characters. All other characters are discarded. The preprocessing process also splits author names in Microsoft Academic into first and last names. In the other data sources, this has already been done by the data provider. In the matching procedure, we treat the first character of the first name of an author as the author's first initial. The preprocessing process also simplifies document titles, source titles, and author names by converting non-US-ASCII characters into US-ASCII characters, for instance by removing accents.

### 3.2. Identification of matching documents

After preprocessing the data, our matching procedure identifies pairs of documents as candidate matches. This is done in six consecutive steps:

1. Matching of documents with the same publication year and DOI.
2. Matching of documents with the same publication year, volume number, and either begin page or article number.
3. Matching of documents with the same publication year, last name of the first author, and either begin page or article number.
4. Matching of documents with the same publication year, last name of the first author, and volume number.
5. Matching of documents with the same publication year, source ID (i.e., ISSN or ISBN), and either begin page or article number.

6. Matching of documents with similar titles. Two documents are considered to have a similar title if the three longest words in the title of the document in Scopus also occur in the title of the document in the other data source.

A matching score is calculated for each pair of documents identified in the above steps as a candidate match. The calculation of the matching score is discussed in Subsection 3.3. A match is established between a pair of documents if the matching score of the documents exceeds a certain threshold. This threshold is set in such a way that the matching procedure favors precision over recall. If a document has a match with multiple other documents, only the match with the highest matching score is considered. When a match between two documents is established in a particular step of the matching procedure, the documents are excluded from the remaining steps of the procedure.

The first step of our matching procedure uses the most restrictive matching criterion. The next steps use less restrictive matching criteria. These criteria yield more candidate matches, making the matching process more demanding from a computational point of view. However, the number of documents that still need to be matched decreases after each step, and in this way the computational cost remains acceptable. In fact, for each of the data sources, at least 80% of the matches are made in the first step of the matching procedure.

Data sources may index multiple versions of (basically) the same document. In some cases, this happens by mistake.[1] In many cases, however, data sources deliberately choose to index multiple versions of basically the same document, for instance a version published in a journal, a version published in a conference proceedings, and a version published in a repository. Our matching procedure creates one-to-one links between documents in Scopus and documents in the other data sources. Suppose for instance that document X is indexed both in Scopus and in Microsoft Academic. Scopus indexes only the version of document X that was published in a journal, while Microsoft Academic indexes also the version that was published in a repository. Most likely, our matching procedure will then create a link between the journal version of document X in Scopus and the journal version of document X in Microsoft Academic. For the repository version of document X in Microsoft Academic, no link will be created. Hence, this version will be seen as part of the unique content of Microsoft Academic relative to Scopus.

---

[1] This problem has been discussed in particular for Scopus. Valderrama-Zurián, Aguilar-Moya, Melero-Fuentes, and Aleixandre-Benavent (2015) and Van Eck and Waltman (2017) found that Scopus sometimes contains multiple records for the same document. More recently, the Scopus team (Baas et al., 2020) reported that the problem of duplicate records has been addressed in a quality improvement program.

**3.3. Calculation of the matching score of two documents**

For a pair of documents identified as a candidate match, a matching score is calculated by comparing the following attributes: (1) DOI, (2) first author (i.e., last name and first initial), (3) document title, (4) source (i.e., ISSN, ISBN, and source title[2]), (5) publication year, (6) volume and issue number, and (7) begin and end page and article number. Each attribute for which there is a match increases the matching score. In the case of the first author, document title, and source title, the matching procedure uses the Levenshtein distance to allow for partial matches. The smaller the Levenshtein distance, the larger the increase in the matching score. A match is established between a pair of documents if the matching score of the documents exceeds a certain threshold.

We refer to the appendix for a more detailed discussion of the calculation of the matching score of two documents.

**3.4. Evaluation**

We evaluated the accuracy of our matching procedure in terms of both recall and precision. We first consider recall, followed by precision.

Recall is the extent to which corresponding documents in two data sources have been matched by our matching procedure. To evaluate the recall of our matching procedure, we manually examined whether non-matched documents in one data source indeed do not have a corresponding document in another data source. Consider for instance the matching of documents in Scopus and CWTS WoS. We first selected all non-matched documents in Scopus and CWTS WoS and we randomly sampled 30 of these documents. For each of the sampled documents, either in Scopus or in CWTS WoS, we then manually tried to identify a corresponding document in the other data source. We took the same approach for the matching of documents in Scopus on the one hand and Dimensions, Crossref, and Microsoft Academic on the other hand. In this way, we considered a total of $4 \times 30 = 120$ non-matched documents. For eight of these documents, we found that our matching procedure had failed to match the document with a corresponding document in the other data source.

---

[2] In our Dimensions data, the titles of conference proceedings are missing. As a consequence, conference proceedings papers that are indexed both in Scopus and in Dimensions may incorrectly not be matched.

Looking in more detail at these eight documents, it turned out that for all of them the correct candidate match had been identified in one of the six steps of our matching procedure (see Subsection 3.2). However, the matching score was below the threshold used by the matching procedure (see Subsection 3.3) and therefore the candidate match had been rejected. The matching score was just below the threshold (i.e., matching score between 25 and 30) in six cases and more substantially below the threshold (i.e., matching score below 25) in the other two cases. The failure of our matching procedure to match two documents was typically caused by inconsistencies in the data provided by two data sources or by incomplete data in one of the data sources. In one of the eight cases, our matching procedure had failed to match two documents, one in Scopus and one in Microsoft Academic, because the Chinese author names and the Chinese document and source titles are presented differently in the two data sources. The author names and the document title are presented in Chinese characters in Microsoft Academic, while in Scopus the author names have been romanized and the document title has been translated to English.

We now turn to precision. Precision is the extent to which the matches made by our matching procedure are correct. To evaluate the precision of our matching procedure, we rely on results of our comparison of the different data sources in terms of the citation links they cover. We manually examined 60 randomly selected citation links found in one data source but not in another. Most of the discrepancies between data sources in the citation links they cover can be expected to be due to incorrect or missing citation links in one of the data sources. However, discrepancies between data sources in the citation links they cover may also be due to documents that have been incorrectly matched by our matching procedure. This applies both to citing documents and to cited documents. For the 60 citing documents and the 60 cited documents that we examined, we found two mistakes made by our matching procedure. We refer to Subsection 5.1 for more details.

## 4. Comparison of coverage of documents

As already mentioned, in our comparison of Scopus, CWTS WoS, Dimensions, Crossref, and Microsoft Academic, we use Scopus as the baseline. Figure 1 shows the differences in coverage of documents between Scopus on the one hand and CWTS WoS, Dimensions, Crossref, and Microsoft Academic on the other hand. Scopus covers 27 million documents. With 23 million documents, CWTS WoS is smaller than Scopus. However, as discussed in Section 2, it is important to keep in mind that two WoS citation indices, the Emerging Sources Citation Index and the Book Citation Index, are not included in CWTS WoS. Dimensions and Crossref are of similar size. They cover respectively 36 and 35 million documents, which is substantially more than Scopus and CWTS WoS. Since Dimensions relies strongly on data from Crossref (Hook et al., 2018), these two data sources largely cover the same documents. Documents covered by Dimensions and not by Crossref typically seem to originate from PubMed. With 73 million documents, Microsoft Academic covers by far the largest number of documents.

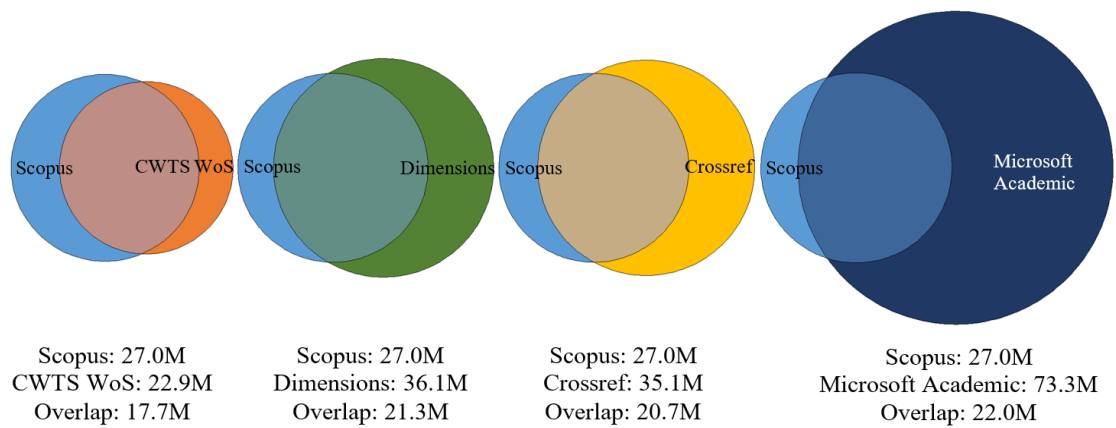

Scopus: 27.0M
CWTS WoS: 22.9M
Overlap: 17.7M

Scopus: 27.0M
Dimensions: 36.1M
Overlap: 21.3M

Scopus: 27.0M
Crossref: 35.1M
Overlap: 20.7M

Scopus: 27.0M
Microsoft Academic: 73.3M
Overlap: 22.0M

Figure 1. Overlap of documents between Scopus and the other data sources.

As can be seen in Figure 1, CWTS WoS has the smallest overlap with Scopus. Almost 18 million documents were found both in CWTS WoS and in Scopus. Dimensions and Crossref each have an overlap of 21 million documents with Scopus. With 22 million documents, Microsoft Academic has the largest overlap with Scopus.

The most striking observation probably is that Microsoft Academic covers so many more documents than the other data sources. Some documents covered by Microsoft Academic are not of a scientific nature. We for instance found news articles and blog posts about someone's private life in Microsoft Academic. To determine the extent to which such non-scientific content artificially inflates the number of documents in Microsoft Academic, we manually examined a random sample of 30 documents that are covered by Microsoft Academic and that do not have a matching document in Scopus. Of these 30 documents, there are four that are clearly not of a scientific nature. The other 26 documents can all be regarded as scientific content. Hence, although Microsoft Academic includes non-scientific content, our manual analysis indicates that this is a small share of the total content of Microsoft Academic. This means that Microsoft Academic provides a much more comprehensive coverage of the scientific literature than the other data sources.

We performed a similar manual examination for a random sample of 30 documents covered by Dimensions and not by Scopus. These documents can all or almost all considered to be of a scientific nature. However, for about one-third of the documents, the scientific contribution does not seem very substantial. These documents include meeting abstracts and other very short items, often with a length of no more than one page. Some of these documents have appeared in journals covered by Scopus, but Scopus has apparently chosen not to index these documents.[3] We made similar observations for a random sample of 30 documents covered by Crossref and not by Scopus. Some documents in Crossref are included in a scientific journal or book, but do not contain any scientific information themselves. In our sample, we for instance found two documents listing the members of the editorial board of a journal. We also found a document containing some of the front matter of a book.

The high-level statistics presented in Figure 1 are of limited value because they hide many important differences between the various data sources. We analyze these differences in the next subsections.

**4.1. Differences in coverage by publication year**

Figure 2 shows the time trend in the number of documents covered by the different data sources and the overlap of documents between Scopus and the other data sources. The yearly number of documents in Dimensions and Crossref is very similar. This illustrates the strong reliance of Dimensions on data from Crossref. The number of documents in Microsoft Academic is substantially smaller in 2017 than in the preceding years. We do not know why this is the case.[4]

---

[3] The Scopus Content Coverage Guide (Scopus, 2020) indicates that Scopus does not cover meeting abstracts and book reviews.

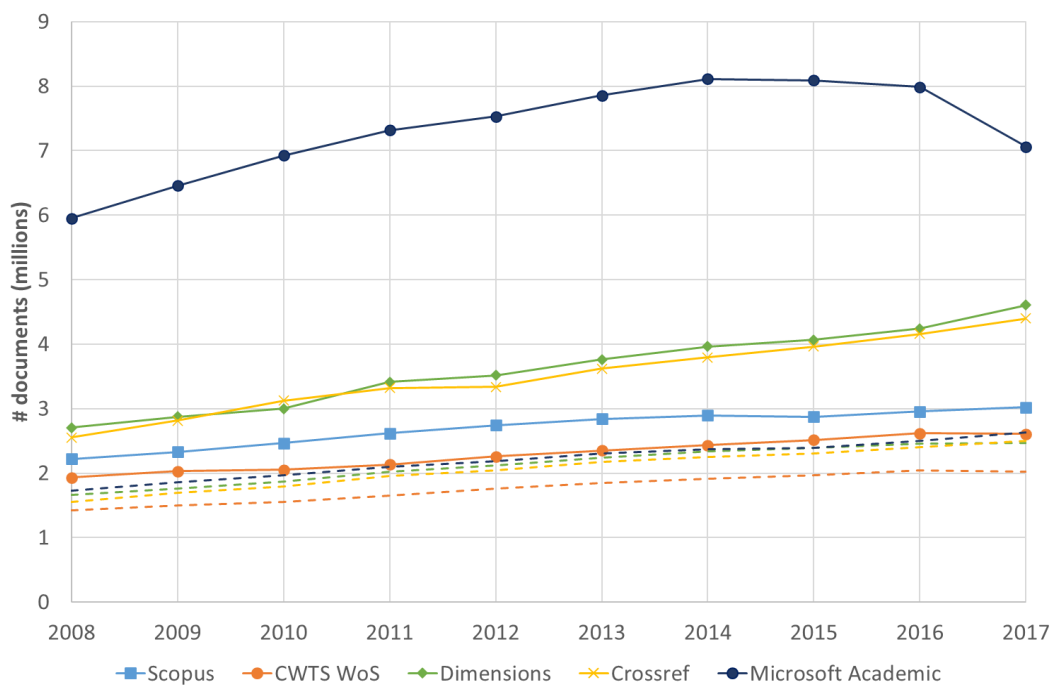

Figure 2. Breakdown by publication year for all documents in each data source (solid line) and for the overlap with Scopus (dashed line).

### 4.2. Differences in coverage by document type

The top-left plot in Figure 3 provides a breakdown by document type for all documents in Scopus and for the overlap with the other data sources. The document type classification of Scopus is used. The plot shows that there are a substantial number of articles and proceedings papers in Scopus for which there are no matching documents in the other data sources. Microsoft Academic has the largest overlap with Scopus, followed by Dimensions and Crossref. CWTS WoS has the smallest overlap with Scopus. It can also be seen that CWTS WoS covers hardly any of the book chapters covered by Scopus. This probably can be explained by the fact that the Book Citation Index is not included in CWTS WoS.

---

[4] It could be that the data for 2017 was not yet complete when the Microsoft Academic data set used in this paper was created. However, this does not seem to be the case. In an analysis of a more recent Microsoft Academic data set, we found a similar drop in the number of documents between 2016 and 2017.

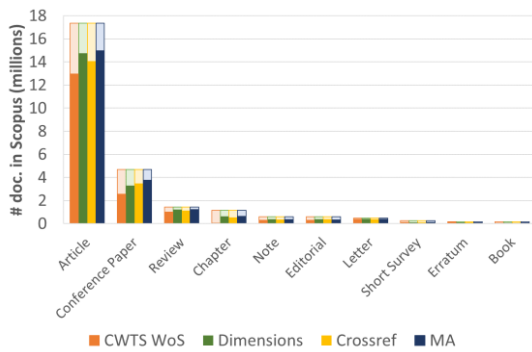

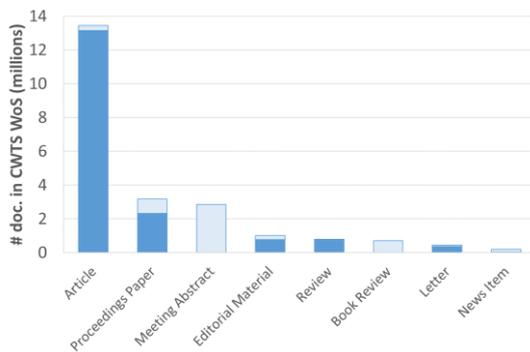
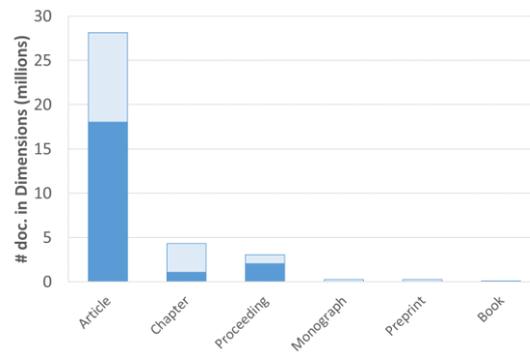

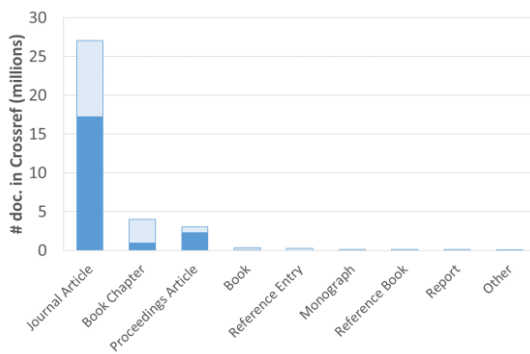
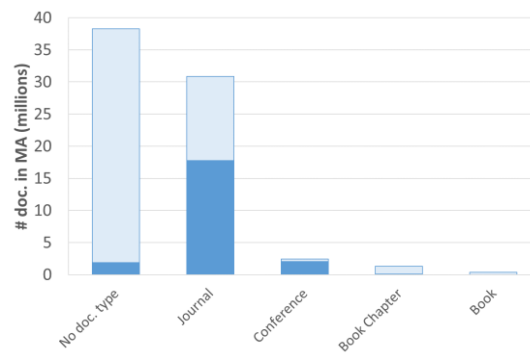

Figure 3. Top-left plot: Breakdown by document type for all documents in Scopus and for the overlap with the other data sources. Other plots: Breakdown by document type for all documents in CWTS WoS, Dimensions, Crossref, and Microsoft Academic and for the overlap with Scopus (in dark blue).

The other plots in Figure 3 provide the opposite perspective. Using the document type classifications of CWTS WoS, Dimensions, Crossref, and Microsoft Academic, these plots offer a breakdown by document type for all documents in CWTS WoS, Dimensions, Crossref, and Microsoft Academic and for the overlap with Scopus. The plot for CWTS WoS shows that meeting abstracts and book reviews are missing in Scopus, which is indeed confirmed by the Scopus Content Coverage Guide (Scopus, 2020). Also, for a substantial number of proceedings papers in CWTS WoS, there are no matching documents in Scopus. On the other hand, almost all articles in CWTS WoS can also be found in Scopus.

Unfortunately, the document type classifications of Dimensions, Crossref, and Microsoft Academic are less detailed. The plots for these data sources therefore offer less information. The plots for Dimensions and Crossref show that for many articles in these data sources there is no matching document in Scopus. Importantly, however, any document published in a journal is classified as an article in Dimensions and Crossref. This may even include content such as the list of editorial board members of a journal or the cover of a journal issue. Dimensions and Crossref also cover many more book chapters than Scopus. Only a small share of the book chapters in Dimensions and Crossref have a matching document in Scopus. For Microsoft Academic, it is hard to draw clear conclusions, since about half of the documents in Microsoft Academic do not have a document type.

**4.3. Differences in coverage by discipline**

We now compare the coverage of documents by broad discipline. In Scopus, documents are assigned to four broad disciplines: *Health Sciences*, *Life Sciences*, *Physical Sciences*, and *Social Sciences & Humanities*. In CWTS WoS, we make use of an assignment of documents to five broad disciplines: *Arts & Humanities*, *Life Sciences & Biomedicine*, *Physical Sciences*, *Social Sciences*, and *Technology*. In Dimensions, we rely on a classification of documents into 22 fields, which we further aggregate into four broad disciplines: *Arts & Humanities*, *Biomedical Sciences*, *Physical Sciences*, and *Social Sciences*. Crossref also provides a classification of documents into broad disciplines, but most documents are not included in this classification. We therefore do not use this classification. We do not use the disciplinary classification of Microsoft Academic either. This classification is missing in the Microsoft Academic data dump that we use.

In the disciplinary classifications of Scopus and CWTS WoS, documents are assigned to disciplines based on the source in which they have appeared. In Scopus, documents in multidisciplinary sources (e.g., *Nature*, *PLOS ONE*, *PNAS*, *Science*, and *Scientific Reports*) are assigned to the *Health Sciences* discipline.[5] In CWTS WoS, these documents do not have an assignment to a discipline. Some documents belong to multiple disciplines in the classifications of Scopus and CWTS WoS. We use a fractional counting approach to handle these documents. We note that in an earlier study significant inaccuracies were identified in the disciplinary classification of Scopus (Wang & Waltman, 2016).

In the disciplinary classification of Dimensions, documents are assigned to disciplines independently of the source in which they have appeared. The accuracy of the disciplinary classification of Dimensions has been questioned (Bornmann, 2018; Herzog & Lunn, 2018; Orduña-Malea & Delgado-López-Cózar, 2018). The classification also has the limitation of being incomplete. Many documents in Dimensions do not have an assignment to a discipline.

The top-left plot in Figure 4 provides a breakdown by discipline for all documents in Scopus and for the overlap with the other data sources. The disciplinary classification of Scopus is used. This for instance means that a document that is covered both by Scopus and by CWTS WoS is assigned to the discipline to which it belongs in the disciplinary classification of Scopus. The disciplinary classification of CWTS WoS plays no role. The plot shows that, in relative terms, the overlap between Scopus and the other data sources is largest in the *Life Sciences* discipline. In the *Social Sciences & Humanities* discipline, the overlap between Scopus and the other data sources, especially CWTS WoS, is quite limited.

The other plots in Figure 4 provide the opposite perspective. Using the disciplinary classifications of CWTS WoS and Dimensions, these plots offer a breakdown by discipline for all documents in CWTS WoS and Dimensions and for the overlap with Scopus. As can be seen in the plot for CWTS WoS, in the *Life Sciences & Biomedicine* discipline, a large number of documents in CWTS WoS do not have matching documents in Scopus. Many of these documents are meeting abstracts, which are not covered by Scopus. From a relative point of view, the large share of the documents in the *Arts & Humanities* discipline in CWTS WoS that do not have matching documents in Scopus is noteworthy. Various types of documents that play a prominent role in the

---

[5] In our Scopus data, documents are assigned to fields, but not to disciplines. To obtain an assignment of documents to disciplines, we use the assignment of fields to disciplines provided at https://service.elsevier.com/app/answers/detail/a_id/15181/supporthub/scopus/.

*Arts & Humanities* discipline in CWTS WoS do not seem to be covered at all by Scopus. The most important one is the WoS document type *Book Review*. Other examples are the WoS document types *Film Review*, *Theater Review*, *Poetry*, and *Fiction, Creative Prose*.

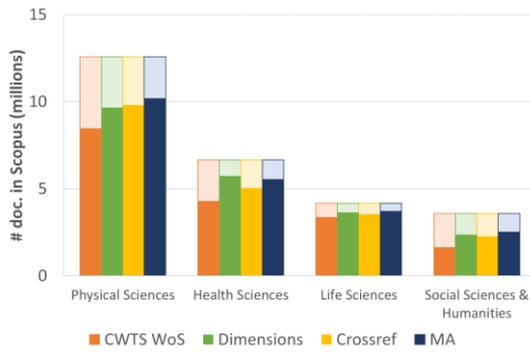

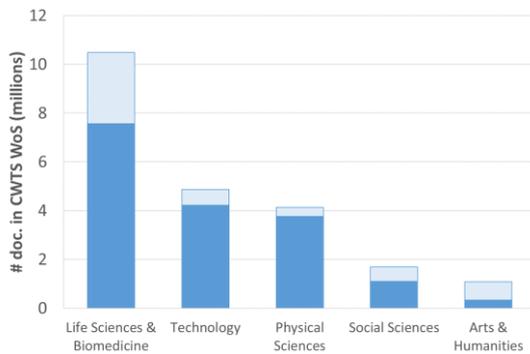
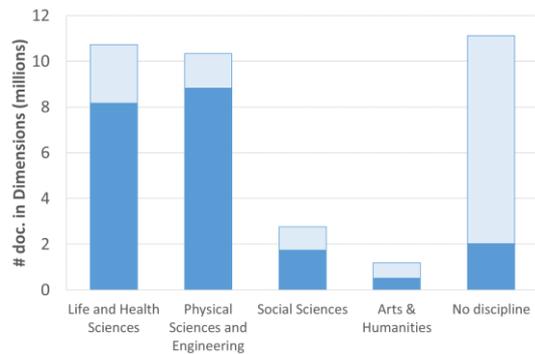

Figure 4. Top-left plot: Breakdown by discipline for all documents in Scopus and for the overlap with the other data sources. Other plots: Breakdown by discipline for all documents in CWTS WoS and Dimensions and for the overlap with Scopus (in dark blue).

The patterns observed for Dimensions are fairly similar to those observed for CWTS WoS. However, one-third of the documents in Dimensions do not have an assignment to a discipline, which limits the conclusions that can be drawn from the results for Dimensions.

**4.4. Differences in coverage by number of references**

The number of references in the reference list of a document may be used as a rough proxy of the scientific contribution of the document. Although there are all kinds of exceptions, a document with many references (e.g., a full research article) may often be considered to make a more substantial scientific contribution than a document with only a few references or no references at all (e.g., an editorial, a letter, or a meeting abstract). For this reason, we look at a breakdown by number of references of the overlap between the different data sources.

The left plot in Figure 5 provides a breakdown by number of references for all documents in Scopus and for the overlap with the other data sources. Documents with a large number of references are overrepresented in the overlap between Scopus and the other data sources. However, even among documents in Scopus with more than 50 references, there are a substantial number for which no matching documents were found in the other data sources.

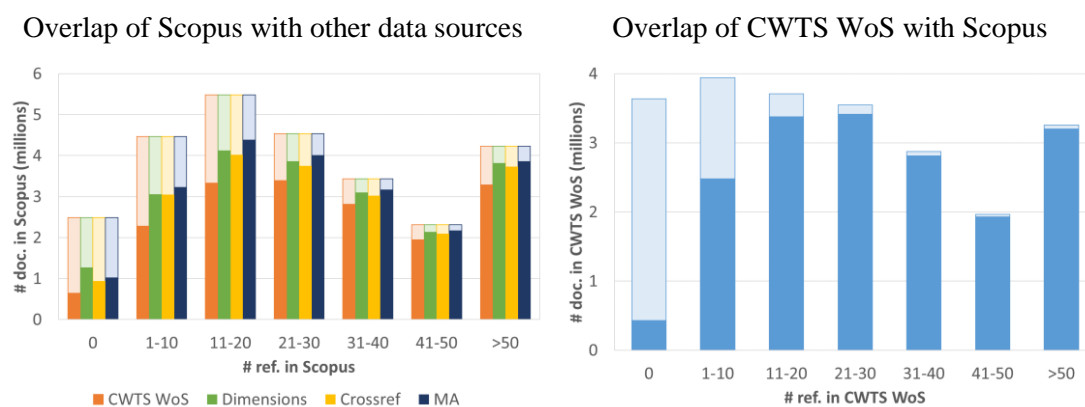

Figure 5. Left plot: Breakdown by number of references for all documents in Scopus and for the overlap with the other data sources. Right plot: Breakdown by number of references for all documents in CWTS WoS and for the overlap with Scopus (in dark blue).

The right plot in Figure 5 offers a breakdown by number of references for all documents in CWTS WoS and for the overlap with Scopus. As can be seen, there are only a very limited number of documents in CWTS WoS that have a large number of references and that do not have a matching document in Scopus.

We do not show results from the viewpoint of Dimensions, Crossref, and Microsoft Academic. In Dimensions and Microsoft Academic, we do not know the total number of references of a document. We know only the number of references that have been matched with a cited document. In Crossref, there are quite a lot of documents for which the reference list is missing because the publisher did not deposit the reference list in Crossref. For these documents, we do not know how many references they have.

**4.5. Differences in coverage by number of citations**

Like the number of references in the reference list of a document, the number of citations received by a document offers a proxy of the scientific contribution of the document. We therefore look at a breakdown by number of citations of the overlap between the different data sources.

The top-left plot in Figure 6 provides a breakdown by number of citations in Scopus for all documents in Scopus and for the overlap with the other data sources. Documents with a larger number of citations are overrepresented in the overlap between Scopus and the other data sources. Almost all documents with more than 25 citations in Scopus have a matching document in the other data sources. A limited number of documents with more than five and no more than 25 citations in Scopus do not have a matching document in the other data sources.

The other plots in Figure 6 provide the opposite perspective. These plots offer a breakdown by number of citations for all documents in CWTS WoS, Dimensions, Crossref, and Microsoft Academic and for the overlap with Scopus. Almost all documents with more than five citations in CWTS WoS have a matching document in Scopus. A limited number of documents with more than five citations in Dimensions, Crossref, and Microsoft Academic do not have a matching document in Scopus.

**4.6. Differences in coverage by language**

Scopus and CWTS WoS are dominated by documents written in English (see also Mongeon & Paul-Hus, 2016). Although they cover a small share of documents written in languages such as Chinese, French, German, Portuguese, and Spanish, 90% of the documents in Scopus and 96% of the documents in CWTS WoS are in English. It is important to keep in mind that the Emerging Sources Citation Index is not included in CWTS WoS. Presumably, this citation index covers a more substantial share of non-English documents. In Dimensions, documents in English are slightly less dominant. 86% of the documents in Dimensions are in English. We do not have language information for Crossref. Likewise, language information is missing in the Microsoft Academic data dump that we use.

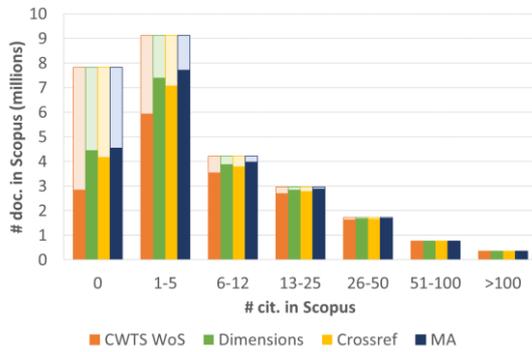

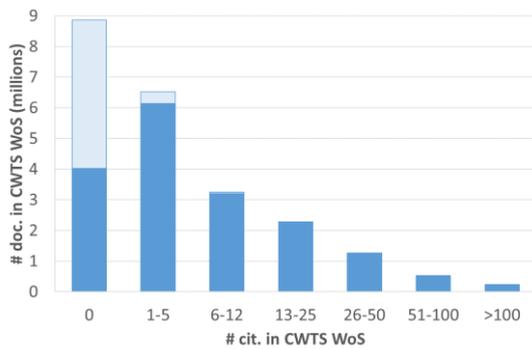
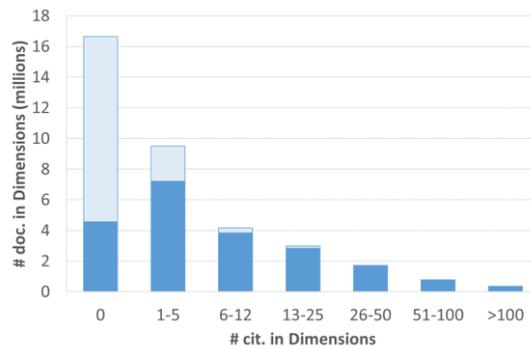

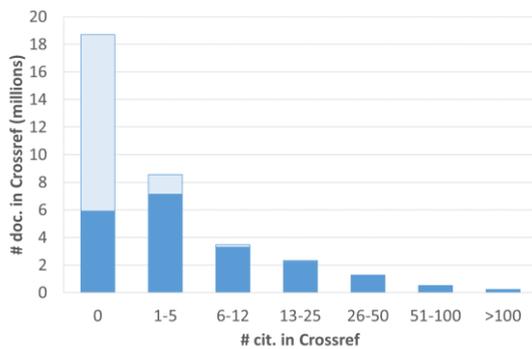
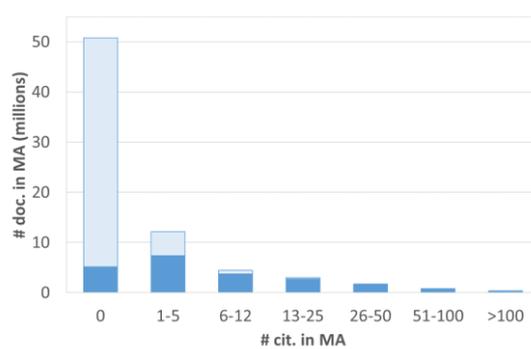

Figure 6. Top-left plot: Breakdown by number of citations for all documents in Scopus and for the overlap with the other data sources. Other plots: Breakdown by number of citations for all documents in CWTS WoS, Dimensions, Crossref, and Microsoft Academic and for the overlap with Scopus (in dark blue).

For most of the documents in Scopus that are not in English, we did not find a matching document in the other data sources. Only about 40% of the non-English documents in Scopus have a matching document in Dimensions or Microsoft Academic, and only 21% have a matching document in CWTS WoS. The other way around, 57% of the non-English documents in CWTS WoS have a matching document in Scopus. In Dimensions, this is the case for only 19% of the non-English documents. These statistics show that Scopus, CWTS WoS, and Dimensions differ a lot in terms of the non-English documents they cover. Although language information is missing in our Microsoft Academic data, this conclusion also extends to Microsoft Academic. In a manual examination of a random sample of 30 documents in Microsoft Academic that do not have a matching document in Scopus, we found that between one-third and half of these documents are not in English.

## 5. Comparison of completeness and accuracy of citation links

To compare the completeness and accuracy of citation links, we again use Scopus as the baseline. We present pairwise comparisons between Scopus on the one hand and CWTS WoS, Dimensions, Crossref, and Microsoft Academic on the other hand. Importantly, in these pairwise comparisons, we consider only citation links between citing and cited documents that are covered by both data sources. Hence, we compare the completeness and accuracy of citation links after correcting for differences in the coverage of documents. The comparisons consider the original citation links made available in the different data sources. They do not consider citation links that may be identified using alternative citation matching algorithms (e.g., Olensky, Schmidt, & Van Eck, 2016).

Figure 7 shows the overlap of citation links between Scopus and the other data sources. Relatively speaking, Scopus and CWTS WoS have the largest overlap. Nevertheless, the discrepancies between the two data sources are quite significant. 1.9% of the citation links in CWTS WoS cannot be found in Scopus. Conversely, 5.8% of the citation links in Scopus cannot be found in CWTS WoS. These discrepancies may be caused by citation links that have been incorrectly identified in Scopus or CWTS WoS. They may also be due to citation links that incorrectly have not been identified in either of these data sources. This will be analyzed below.

The discrepancies between Scopus on the one hand and Dimensions and Microsoft Academic on the other hand are even larger. 3.4% of the citation links in Dimensions cannot be found in Scopus. Moreover, for 10.6% of the citation links in Scopus, there is no corresponding citation link in Dimensions. Likewise, 5.1% of the citation links in Microsoft Academic cannot be found in Scopus, while 12.7% of the citation links in Scopus do not have a corresponding citation link in Microsoft Academic.

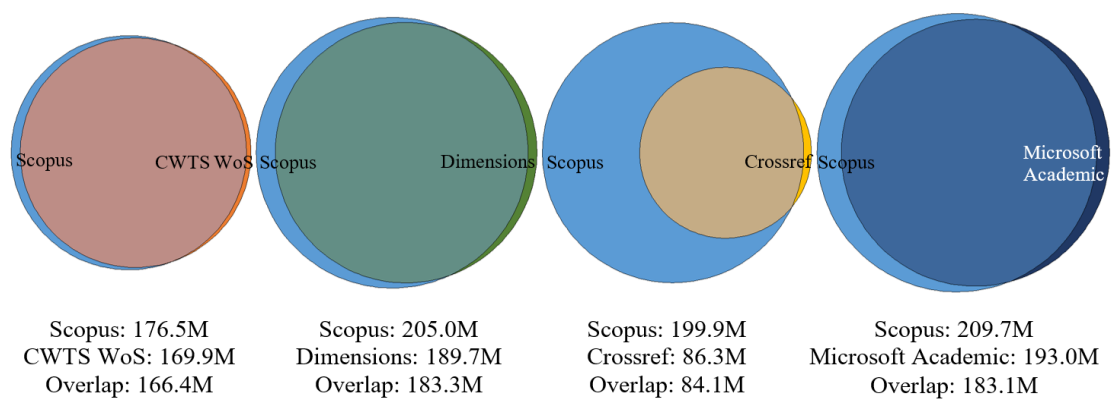

Scopus: 176.5M           Scopus: 205.0M            Scopus: 199.9M            Scopus: 209.7M
CWTS WoS: 169.9M      Dimensions: 189.7M      Crossref: 86.3M            Microsoft Academic: 193.0M
Overlap: 166.4M          Overlap: 183.3M            Overlap: 84.1M              Overlap: 183.1M

Figure 7. Overlap of citation links between Scopus and the other data sources.

Finally, comparing Scopus and Crossref, we find that 57.9% of the citation links in Scopus cannot be obtained from Crossref. There are three main reasons for this. First, some publishers deposit documents in Crossref without depositing their references. Second, there are publishers (in particular ACS, Elsevier, and IEEE) that deposit references in Crossref but choose not to make these references openly available.[6] Third, Crossref has suffered from a technical problem due to which a large number of openly available references incorrectly have not been linked to cited documents (Bilder, 2019).

Figure 7 makes clear that Dimensions has an important advantage over Crossref. Our earlier results indicate that Dimensions and Crossref have a fairly similar coverage of documents, but Figure 7 shows that Dimensions provides access to many more citation links than Crossref. Although Dimensions relies strongly on data from Crossref, it also benefits from data received directly from publishers, enabling the Crossref data to be enriched in various ways, in particular by adding citation links, but also by adding abstracts, affiliation data, and so on.

---

[6] When we performed our analysis, Elsevier did not yet make its references openly available in Crossref. In January 2021, Elsevier will open its references (Plume, 2020). This will result in a large increase in the number of citation links that can be obtained from Crossref.

**5.1. Analysis of incompleteness or inaccuracy of citation links**

The discrepancies shown in Figure 7 between the different data sources are quite significant. To better understand these discrepancies, we now analyze the incompleteness or inaccuracy of citation links in the various data sources.

An important explanation for the discrepancies in the citation links covered by the various data sources is that for some documents no reference list is available in some of the data sources. Missing reference lists are an important explanation for citation links in Scopus for which there is no corresponding citation link in Dimensions, Crossref, or Microsoft Academic. For 15 million citation links in Scopus, the citing document does not have a reference list in Dimensions. Likewise, there are 18 million citation links in Scopus for which the citing document does not have a reference list in Microsoft Academic. In Crossref, missing reference lists are a major problem. Missing reference lists in Crossref are responsible for 107 million of the 116 million citation links in Scopus for which there is no corresponding citation link in Crossref. Of these 107 million citation links, 27 million are due to reference lists that have not been deposited in Crossref at all and 80 million are due to reference lists that have been deposited but that the publisher has chosen not to make openly available. In CWTS WoS, missing reference lists are highly exceptional. Of the 10 million citation links in Scopus for which there is no corresponding citation link in CWTS WoS, only 0.1 million are due to missing reference lists in CWTS WoS. Finally, in Scopus, the problem of missing reference lists is more significant than in CWTS WoS but less serious than in the other data sources. CWTS WoS, Dimensions, Crossref, and Microsoft Academic each cover between 1 and 2 million citation links for which the citing document does not have a reference list in Scopus.

In earlier work (Van Eck & Waltman, 2017; see also Olensky et al., 2016), we studied inaccuracies of citation links in Scopus and WoS. For WoS, three problems were identified. First, some references are missing in the reference lists of documents in WoS. Second, sometimes there is an error in a reference in WoS, such as an incorrect publication year or volume number. Third, some references in WoS have been incorrectly matched with a cited document, leading to so-called phantom citations (García-Pérez, 2010). For Scopus, the opposite problem was identified. Some references incorrectly have not been matched with a cited document, even though all information needed to make a match seems to be available.

We now look in more detail at the discrepancies in the citation links covered by Scopus on the one hand and Dimensions and Microsoft Academic on the other hand, focusing on discrepancies that are not due to documents for which no reference list is available. We manually examined 15 randomly selected citation links in Scopus that are not in Dimensions and 15 randomly selected citation links in Scopus that are not in Microsoft Academic. It turns out that in about two-third of the cases Dimensions or Microsoft Academic incorrectly has not identified a citation link. Hence, these data sources both fail to identify a substantial number of citation links. We found just a few cases in which a citation link has been incorrectly identified in Scopus.

The other way around, we also performed a manual examination of 15 randomly selected citation links in Dimensions that are not in Scopus and 15 randomly selected citation links in Microsoft Academic that are not in Scopus. Of the citation links in Dimensions that are not in Scopus, about half incorrectly have not been identified in Scopus. A few citation links have been incorrectly identified in Dimensions. Of the citation links in Microsoft Academic that are not in Scopus, only one incorrectly has not been identified in Scopus. About one-third of the citation links have been incorrectly identified in Microsoft Academic. We also found a substantial number of cases in which Scopus and Microsoft Academic seem to make different choices, causing a citation link to be created in Microsoft Academic but not in Scopus. Some cases involve in-print references (i.e., references to a document that has not yet formally been published), for which Microsoft Academic tries to create a citation link, while Scopus does not seem to do so. Other cases involve references to 'secondary' versions of a document (i.e., references to for instance a preprint or a proceedings paper instead of a journal article). For such references, it seems that Microsoft Academic chooses to create a citation link to the 'primary' version of the document (usually a journal article), while Scopus does not do so.[7]

---

[7] Two comments regarding our manual examination of citation links are in order. First, there were a few citation links for which a full examination was not possible because the citing document was behind a paywall and we did not have access to the reference list of this document. For these citation links, we were unable to determine whether they have been correctly identified or not. Second, Scopus reports a precision of 99.9% and a recall of 98.3% for its citation matching algorithm (Baas et al., 2020). Our findings seem to confirm the high precision and recall of citation links in Scopus, although the precision seems to be lower than reported by Scopus.

In total, we manually examined 60 citation links that can be found in one data source but not in another. In only two cases, we found that the discrepancy is due to a mistake made by our procedure for matching documents in Scopus with documents in the other data sources (see Section 3). Hence, in a sample of 60 citing documents and 60 cited documents, we found only two mistakes made by our matching procedure. This indicates that the matching procedure is sufficiently precise.

## 6. Conclusions

The value of a bibliographic data source depends on many different elements. The coverage of a data source is very important, but the completeness and accuracy of the data provided by a data source are of course important as well. For some purposes, the speed of updating is also a key concern. Another crucial issue for determining the value of a bibliographic data source is the way in which the data is made available, for instance through web interfaces, APIs, and data dumps. Finally, the conditions under which a data source can be used are of major importance (Waltman & Larivière, 2020).

While we recognize the importance of all these elements, we have chosen a specific focus for the analysis presented in this paper. In our comparison of Scopus, CWTS WoS, Dimensions, Crossref, and Microsoft Academic, our focus has been on differences between the data sources in the coverage of documents and in the completeness and accuracy of citation links. In addition, we have chosen to consider only scientific documents in our analysis. Some data sources, in particular Dimensions, also provide data on other types of entities, but this data falls outside the scope of our analysis.

The main findings of our analysis can be summarized as follows:

- Comparing Scopus and CWTS WoS, it turns out that Scopus covers a large number of documents that are not covered by CWTS WoS, including documents with substantial numbers of references and citations. Documents covered by Scopus and not by CWTS WoS have appeared mostly in journals and conference proceedings. We have also identified a substantial number of book chapters covered by Scopus and not by CWTS WoS, but this is likely to be a consequence of the fact that the Book Citation Index is not included in CWTS WoS. Almost all journal articles covered by CWTS WoS are also covered by Scopus. However, CWTS WoS covers meeting abstracts and book reviews,

- which are not covered by Scopus. A substantial share of the proceedings papers covered by CWTS WoS are not covered by Scopus either.
- The results of the comparison of Scopus with Dimensions and Crossref are somewhat more difficult to interpret. This is partly due to limitations of the document type classifications of Dimensions and Crossref. These classifications do not distinguish between different types of documents published in journals. Dimensions and Crossref turn out to have a similar coverage of documents. This illustrates the strong reliance of Dimensions on data from Crossref.

  Scopus covers a large number of journal articles that are not covered by Dimensions and Crossref. The other way around, Dimensions and Crossref cover an even larger number of documents that have been published in journals and that are not covered by Scopus. However, a significant share of these documents are meeting abstracts and other short items that do not seem to make a very substantial scientific contribution. Dimensions and Crossref also cover many book chapters and quite some proceedings papers that are not covered by Scopus. On the other hand, Scopus also covers many proceedings papers that are not covered by Dimensions and Crossref.
- Of the five data sources studied in this paper, Microsoft Academic offers by far the most comprehensive coverage of the scientific literature. It covers many more documents than the other data sources. Microsoft Academic provides only a very basic document type classification, which does not give much insight into the nature of the documents covered by Microsoft Academic. However, a manual examination of a sample of documents covered by Microsoft Academic and not by Scopus has confirmed that most of these documents are indeed of a scientific nature. It has also shown that Microsoft Academic covers many documents that are not in English.

  Despite the large coverage of Microsoft Academic, there are still quite a lot of documents in Scopus without a matching document in Microsoft Academic. This includes journal articles and also proceedings papers and book chapters.
- All data sources suffer from problems of incompleteness and inaccuracy of citation links. However, our overall conclusion is that, in terms of the quality of citation links, the more established data sources, Scopus and CWTS WoS, outperform two recent alternatives, Dimensions and Microsoft Academic.

Missing citation links are a significant problem in Dimensions and Microsoft Academic. These data sources also have the limitation that they do not provide data for references that have not been matched with a cited document. In the case of CWTS WoS, we are especially concerned about the problem of phantom citations (García-Pérez, 2010; Van Eck & Waltman, 2017).

In Crossref, incompleteness of citation links is a major problem. This is partly caused by publishers that do not deposit references in Crossref. To a significant extent, however, this is due to publishers that do deposit references in Crossref but choose not to make these references openly available. Citation links resulting from closed references are available within Crossref's internal infrastructure, but they are not accessible to the outside world. Crossref takes these closed citation links into account in the aggregate citation counts it provides for documents (Heibi, Peroni, & Shotton, 2019a). This for instance explains why Harzing (2019) concludes that Crossref has "a similar or better coverage" of citations than Scopus and WoS. However, while closed citation links are taken into account in aggregate citation counts provided by Crossref, the individual citation links cannot be accessed.

How the differences between the data sources should be assessed depends on the purpose for which the data sources are used. For many purposes, a broad coverage of documents is valuable, for instance to make sure that locally relevant research is properly taken into account (e.g., Hicks, Wouters, Waltman, De Rijcke, & Rafols, 2015) and to obtain a good coverage of the literature in disciplines in which researchers prefer to publish proceedings papers or books rather than journal articles. However, for other purposes, it may be desirable to work within a more restricted universe of documents (e.g., López-Illescas, de Moya Anegón, & Moed, 2009). For instance, to enable meaningful international comparisons of universities, documents that have not been published in international scientific journals are deliberately excluded from the calculation of the bibliometric statistics reported in the CWTS Leiden Ranking ([www.leidenranking.com](www.leidenranking.com)).

In our view, there is value both in the comprehensiveness offered by Dimensions and Microsoft Academic and in the selectivity offered by Scopus and WoS. However, comprehensiveness and selectivity do not need to be seen as mutually exclusive. In line with the philosophy of the developers of Dimensions (Herzog et al., 2020; Hook et al., 2018), we believe that data sources should be as comprehensive as possible while filters for making relevant selections of the scientific literature should be provided on top of the data. Depending on the purpose for which a data source is used, one may or may not wish to apply certain filters to restrict an analysis to a particular selection of the scientific literature. In this approach, comprehensiveness and selectivity are no longer mutually exclusive. The ideal data source provides a comprehensive coverage of the scientific literature, like Dimensions and Microsoft Academic already aim to do, and in addition it also offers a flexible set of filters for making selections of the literature. Important examples of such filters are expert-curated journal lists, such as those provided by Scopus, WoS, Directory of Open Access Journals, and many others. The fine-grained document type classifications of Scopus and WoS offer another example.

## 7. Limitations

Our work has several limitations. First of all, our analysis is not entirely up-to-date, since it is based on data sets from 2018 and 2019. The data sources studied in this paper are regularly being improved and expanded. The most recent developments are not covered by our analysis, and some of our findings may therefore not be fully representative for the current state of the different data sources. Furthermore, we have performed pairwise comparisons between Scopus and the other data sources. CWTS WoS, Dimensions, Crossref, and Microsoft Academic have not been compared directly with each other. In addition, in the case of CWTS WoS, two citation indices that are part of the WoS Core Collection, the Emerging Sources Citation Index and the Book Citation Index, are not included. This is an important limitation that needs to be kept in mind in the interpretation of our findings for WoS. Finally, our procedure for matching documents in Scopus with documents in the other data sources is somewhat conservative. Avoiding false positives (i.e., documents that have been incorrectly matched) is considered more important than avoiding false negatives (i.e., documents that incorrectly have not been matched). This means that our analysis underestimates the true overlap between Scopus and the other data sources.

## Acknowledgements

We would like to thank Susanne Steiginga, Wim Meester, and Michiel Schotten (Scopus, Elsevier), Jonathan Adams (Web of Science, Clarivate Analytics), Christian Herzog (Dimensions, Digital Science), and Kuansan Wang (Microsoft Academic),

Alberto Martín-Martín, and an anonymous reviewer for their helpful comments on earlier versions of this paper.

## Author contributions

Martijn Visser: conceptualization, data curation, formal analysis, investigation, methodology, software, visualization, writing - review & editing

Nees Jan van Eck: conceptualization, data curation, methodology, visualization, writing - review & editing

Ludo Waltman: conceptualization, investigation, methodology, writing - original draft

## Competing interests

The authors are affiliated with the Centre for Science and Technology Studies (CWTS) at Leiden University. CWTS has commercial relationships with Clarivate Analytics, Elsevier, and Digital Science, the producers of Web of Science, Scopus, and Dimensions, respectively. CWTS and Digital Science also work together as founding partners of the Research on Research Institute (RoRI; https://researchonresearch.org). Waltman works together with OpenCitations in his capacity as chair of the Advisory Board of the Research Centre for Open Scholarly Metadata.

## Data availability

The Scopus and Dimensions data used in this paper has been made freely available to CWTS for research purposes. The Web of Science data has been made available to CWTS under a paid license. For Crossref, we made use of openly available data obtained through the Crossref API. For Microsoft Academic, we made use of an openly available data dump (Microsoft Academic, 2019). We are not allowed to redistribute the Scopus, Web of Science, and Dimensions data used in this paper. The data therefore cannot be made available. The statistics presented in the figures in this paper are available in Zenodo (Visser, Van Eck, & Waltman, 2020).

## References

Baas, J., Schotten, M., Plume, A., Côté, G., & Karimi, R. (2020). Scopus as a curated, high-quality bibliometric data source for academic research in quantitative science

**Appendix: Calculation of the matching score of two documents**

In this appendix, we provide a detailed discussion of the approach taken in our matching procedure (see Section 3) to calculate the matching score of two documents. There are many ways in which a matching score could be calculated. We performed extensive comparisons of different approaches to calculate matching scores, resulting in the approach discussed below.

Consider two documents, A and B, for which we want to calculate the matching score. Document A is a document in Scopus. Document B is a document in CWTS WoS, Dimensions, Crossref, or Microsoft Academic. We first determine the extent to which the documents have a match based on DOI, first author, document title, source, and other attributes. By combining the results obtained for each of these attributes, we then determine the matching score of the documents.

Documents A and B have a match based on DOI if they have exactly the same DOI. Hence, $m_{\text{DOI}}$ equals 1 if the DOIs of documents A and B are identical and 0 otherwise.

The extent to which documents A and B have a match based on first author is determined by the similarity of the first authors of the documents in terms of their last name and their first initial. More specifically,

$$m_{\text{first author}} = 0.8 - 0.8\, D(l_A, l_B)/\max(L(l_A), L(l_B)) + 0.2 E(f_A, f_B),$$

where $l_X$ denotes the last name of the first author of document X and $f_X$ denotes the first initial of the first author of document X.[8] $L(a)$ equals the length of string $a$. $D(a,b)$ equals the Levenshtein distance between string $a$ and string $b$. $E(a,b)$ equals 1 if $a$ and $b$ are identical and 0 otherwise. It follows from the above equation that $m_{\text{first author}}$ equals 1 if the first authors of documents A and B have the same last name and the same first initial.

The extent to which documents A and B have a match based on title is determined by the similarity of their titles. More specifically,

$$m_{\text{title}} = 1 - D(t_A, t_B)/\max(L(t_A), L(t_B)),$$

where $t_X$ denotes the title of document X. $m_{\text{title}}$ equals 1 if the titles of documents A and B are identical.

The extent to which documents A and B have a match based on source is determined by looking at ISSNs, ISBNs, and source titles. $m_{\text{source}}$ equals 1 if the source of document A has the same ISSN or ISBN as the source of document B. If the ISSN or ISBN is not the same, $m_{\text{source}}$ is given by

$$m_{\text{source}} = 1 - [D(s_A, s_B) - |L(s_A) - L(s_B)|]/\min(L(s_A), L(s_B)),$$

where $s_X$ denotes the title of the source of document X. $m_{\text{source}}$ equals 1 if the title of the source of document A is contained within the title of the source of document B or the other way around. There may be multiple variants of the title of a source, for instance a full variant (e.g., 'Journal of the Association for Information Science and Technology') and an abbreviated one (e.g., 'J. Assoc. Inf. Sci. Technol.'). In that case, $m_{\text{source}}$ is calculated for each combination of a source title variant for document A and a source title variant for document B and the highest value of $m_{\text{source}}$ is used.

---

[8] In the final preparation of this paper, we found a small mistake in our code for calculating $m_{\text{first author}}$. In our code, $m_{\text{first author}}$ is incorrectly calculated as $m_{\text{first author}} = 0.8 - D(l_A, l_B)/\max(L(l_A), L(l_B)) + 0.2 E(f_A, f_B)$, resulting in values that are somewhat lower than intended. The consequences of this mistake are very limited. No significant effects on the results of our analysis are to be expected.

Finally, the extent to which documents A and B have a match based on other attributes is given by

$$m_{\text{other}} = 0.1E(y_A, y_B) + 0.2E(v_A, v_B) + 0.1E(i_A, i_B) + 0.3E(b_A, b_B) + 0.3E(e_A, e_B),$$

where $y_X$, $v_X$, $i_X$, $b_X$, and $e_X$ denote, respectively, the publication year, the volume number, the issue number, the begin page, and the end page of document X. $m_{\text{other}}$ equals 1 if all five attributes are identical for documents A and B. Instead of a match based on begin page, we also allow for a match based on article number.

Based on $m_{\text{DOI}}$, $m_{\text{first author}}$, $m_{\text{title}}$, $m_{\text{source}}$, and $m_{\text{other}}$, the matching score of documents A and B is given by

$$S_{A,B} = 15m_{\text{DOI}} + 7m_{\text{first author}} + 14m_{\text{title}} + 5m_{\text{source}} + 14m_{\text{other}}.$$

To establish a match between documents A and B, $S_{A,B}$ must be greater than 30. Using a threshold of 30, our matching procedure (see Section 3) is relatively conservative and favors precision over recall. The matching procedure can be made less conservative by decreasing the threshold. The results produced by the matching procedure are relatively insensitive to a decrease in the threshold. For CWTS WoS, Dimensions, and Crossref, decreasing the threshold from 30 to 25 leads to an increase of less than 2% in the number of documents for which a matching document is identified in Scopus. For Microsoft Academic, the increase is less than 4%.